\documentstyle[11pt]{article}

\newcommand{\nc}{\newcommand}
\nc{\be}{\begin{equation}}
\nc{\ee}{\end{equation}}
\nc{\bea}{\begin{eqnarray}}
\nc{\eea}{\end{eqnarray}}
\nc{\RR} {\rangle }
\nc{\LL} {\langle }
\nc{\rmi}[1]{{\mbox{\small #1}}}
\nc{\eq}[1]{eq.~(\ref{#1})}
\nc{\Eq}[1]{Eq.~(\ref{#1})}
\nc{\ul}{\underline}
\nc{\mc}{\multicolumn}
\nc{\h}{{\cal H}}
\nc{\e}   {\exp }
\nc{\sh}  {\sinh }
\nc{\ch}  {\cosh }
\nc{\th}  {\tanh }
\nc{\cth} {\coth }
\nc{\half}{\mbox{\small$\frac12$}}
\nc{\eights}{\mbox{\small$\frac18$}}
\begin{document}

\begin{titlepage}

\vskip0.5cm

\begin{flushright}
MS-TPI-96-2 \\
DFTT 8/96   \\
February 1996
\end{flushright}
\vskip0.5cm

{\Large\bf
\begin{center} On the Universality of Certain \\ Non-Renormalizable
Contributions \\ in Two-Dimensional Quantum Field Theory
\end{center}
}
\vskip1.3cm
\centerline{M. Caselle$\,{}^a$ and K. Pinn$\,{}^{b}$}
\vskip5mm
\centerline{\sl $^a$ Dipartimento di Fisica Teorica dell'Universit\`a di
Torino,}
\centerline{\sl Istituto Nazionale di Fisica Nucleare, Sezione di Torino}
\centerline{\sl Via P. Giura 1, I-10125 Torino, Italy
\footnote{e--mail: caselle~@to.infn.it}}
\vskip2mm
\centerline{\sl $^b$ Institut f\"ur Theoretische Physik I, Universit\"at
                M\"unster}
\centerline{\sl Wilhelm-Klemm-Str.\ 9, D-48149 M\"unster, Germany
\footnote{e--mail: pinn~@uni-muenster.de}}

\vskip2.0cm

\begin{abstract}
We consider the question of removing the ultraviolet cutoff  in a 2D
Quantum Field Theory with an interaction term which is
non-renormalizable by power counting.  This model arises as the first
non-trivial correction  beyond the Gaussian approximation of the so
called  Capillary Wave or Drumhead Model,
and is rather important from a physical
point of view since it correctly describes the finite size effects of
two-dimensional interfaces. Despite the fact that the interaction is
non-renormalizable,  we prove that for a large class of regularization
schemes the  finite and divergent parts can be separated in a simple
way.  Furthermore, the finite part is independent of the  choice of
cutoff prescription used.
\end{abstract}

\end{titlepage}

\setcounter{footnote}{0}
\def\thefootnote{\arabic{footnote}}
\section{Introduction}

A model which is widely used to describe fluid interfaces is the so
called Capillary Wave Model (CWM)~\cite{cwm}. In the literature, this
model is also known as Drumhead Model. It is derived from the assumption
that the fluctuations of the interface are described by an effective
Hamiltonian proportional to the variation of the surface area with
respect to the equilibrium solution (which is the solution which
minimizes the area). In the case of an interface bounded by a rectangle
of size $L_1\times L_2$ the Hamiltonian has the following form:
\be\label{cwm}
\h = \sigma \int_0^{L_1} dx_1 \int_0^{L_2} dx_2 \, \,
\sqrt{ 1 + [\nabla \tilde\phi(x_1,x_2)]^2} \, ,
\ee
where $\sigma$ denotes the interface tension, and the field
$\tilde\phi(x_1,x_2)$ describes the interface displacement from the
equilibrium position as a function of the two longitudinal coordinates
$x_1$ and $x_2$. We shall adopt in the following  periodic boundary
conditions,  i.e.\ $\tilde\phi(x_1,x_2)=\tilde\phi(x_1+L_1,x_2)$ and
$\tilde\phi(x_1,x_2)= \tilde\phi(x_1,x_2+L_2)$.  It is possible to
generalize  our results to other boundary conditions.

Due to its non-polynomial nature, until recently the CWM has been
studied only in its quadratic approximation. In this limit the
Hamiltonian (\ref{cwm}) becomes equivalent to a massless 2D Gaussian
model, see~\cite{thegang} for details. Recently, in~\cite{thegang,pv}
the first correction beyond the Gaussian approximation was  studied in
detail. Despite the fact that this corresponds to a scalar field theory
with a non-renormalizable interaction term we found strong evidence
that the resulting contribution to the interface free energy is
universal, i.e.\ independent of the regularization scheme. We gave some
general arguments supporting this assertion and also tested two different
regularization schemes, finding the same cutoff independent
contribution. In this paper we want to further pursue this analysis, by
studying a wider class  of regularization schemes.

Let us mention two more reasons to be interested   in this model. First,
it can be shown by duality, that the Hamiltonian~(\ref{cwm}) can be used
to describe the physics of Wilson loops in the confining regime of
Yang-Mills theories. In particular, the model describes Wilson loops
beyond the roughening transition, hence in the region (which is the most
interesting from a physical point of view) in which the continuum limit
can be taken. In this context the Gaussian approximation of the CWM
Hamiltonian was discussed for the first time by  L\"uscher, Symanzik and
Weisz in~\cite{lsw}.  Second, eq.~(\ref{cwm}) coincides with the
Nambu-Goto model for the bosonic string (in a special frame).   The
corresponding quantum theory is anomalous. Depending on the quantization
method one finds either the breaking of rotational invariance or the
appearance of interacting longitudinal modes. However, it can be
shown~\cite{olesen} that in the infrared limit,  i.e.\ when $L_1$ and
$L_2$ become large, these anomalous features disappear and the string
theory  becomes a simple Conformal Field Theory, with a well defined
quantum behavior. This limit exactly coincides with  the Gaussian
approximation mentioned above.

This paper is organized as follows: In section~2 we discuss the
expansion of the CWM at the first order beyond the Gaussian
approximation, to be called 2-loop contribution in the following. We
claim that in the ultraviolet limit a finite part of this contribution
can be extracted in a natural way and is independent of the choice of
the regularization scheme.  This claim is made precise in form of a
statement at the end  of section~2.  In section~3 we explicitly compute
the cutoff dependence of the  2-loop contribution for a continuous
momentum space cutoff  and for the lattice cutoff scheme,  thus
providing proofs of our statement for these two important regularization
schemes.

\section{2-Loop Expansion of the CWM}

Our aim is to evaluate the partition function
\be\label{parti}
Z = \int D \tilde\phi \,
\exp \left\{
- \sigma \int_0^{L_1} dx_1 \int_0^{L_2} dx_2 \,
\sqrt{ 1 + [\nabla \tilde\phi(x_1,x_2)]^2} \,
 \right\} \, .
\ee
We rescale  $\tilde\phi(x_1,x_2)=\phi(\xi_1,\xi_2)/\sqrt{\sigma}$,
where $x_i = \xi_i \, L_i$.
Then we expand in the adimensional variable
$(\sigma L_1L_2)^{-1}$.
The Hamiltonian~(\ref{cwm}) becomes
\be\label{Atot}
\h\left[ \phi\right] = \sigma L_1L_2 +
\h_g\left[ \phi\right] -\frac{1}{\sigma \, L_1 L_2 }
\h_{p}\left[ \phi\right] + O\Bigl((\sigma L_1 L_2)^{-2}\Bigr)  \, ,
\ee
with
\bea
\h_g\left[ \phi\right]&=&
 \half
\int_0^1 d\xi_1 \int_0^1 d\xi_2 \, [\nabla_u \phi(\xi_1,\xi_2)]^2
 \label{Ag}\\
\h_{p}\left[ \phi\right]&=& \eights
\int_0^1 d\xi_1 \int_0^1 d\xi_2 \,
\Bigl([\nabla_u \phi(\xi_1,\xi_2)]^2\Bigr)^2
\label{Ap}~~.
\eea
The `asymmetric gradient' is defined through
\be
\nabla_u = \left( u^{1/2} \frac{\partial}{\partial \xi_1} \, , \,
u^{-1/2} \frac{\partial}{\partial \xi_2}
\right) \equiv \left( \nabla_{u,1},\nabla_{u,2} \right)
 \,  \mbox{\ with \ } u\equiv L_2/L_1 \, .
\ee
The corresponding (formal) expansion
of the partition function~\eq{parti} is
\be
Z= {\rm const} \cdot \exp(-\sigma L_1 L_2) \cdot Z_1 \cdot
\left( 1 + \frac{1}{8 \sigma L_1 L_2}
\int_0^{1} d\xi_1 \int_0^{1} d \xi_2
 \left\langle \left(
[\nabla_u \phi(\xi_1,\xi_2)]^2
\right)^2 \right\rangle + \dots \right) \, .
\ee
Here, we have introduced the Gaussian measure
\be
\langle {\cal O} \rangle \equiv \frac{1}{Z_1}
\int_{a} D \phi \, \exp\left( - \half
\int_0^1 d \xi_1 \int_0^1 d \xi_2 \, [\nabla_u \phi(\xi_1,\xi_2)]^2 \right)
\, {\cal O}(\phi) \, .
\ee
The first order (1-loop) contribution $Z_1$ to \eq{parti} is
\be\label{1loop}
Z_1 =
\int_{a} D \phi \, \exp\left( - \half
\int_0^1 d\xi_1 \int_0^1 d\xi_2 \, [\nabla_u \phi(\xi_1,\xi_2)]^2 \right)
\, .
\ee
The subscript $a$ indicates that the functional integral (which is
divergent in the ultraviolet) is to be regularized by a suitable cutoff.
Examples for cutoff schemes will  be given below.  $Z_1$  is a purely
Gaussian integral and can be evaluated by using standard techniques. The
result turns out to be a function of the adimensional ratio $u$ alone:
\be
Z_1 \left(u \right) = \frac{1}{\sqrt{u}}
\Bigl| \eta\left(i u\right)/\eta\left(i\right) \Bigr|^{-2}~~,
\label{Zg}
\ee
where $\eta$ is the Dedekind eta function
\be
\eta(\tau)=q^{1/24}\prod_{n=1}^{\infty}\left(1-q^n\right)~~,
\quad\quad q\equiv \exp(2\pi i \tau)~~.
\label{eta}
\ee
This is a well known result in string theory and conformal field
theory, and coincides\footnote{The constant $\eta\left(i \right)$
has been introduced to normalize $Z_1(1)=1$.}
with the partition function of a 2D conformal invariant free
boson on a torus of modular parameter $\tau=i u $ \cite{cgv}.

The calculation of the 2-loop contribution is less simple.
One can show that
\be\label{diff}
\left.
 \left\langle \left(
[\nabla_u \phi(\xi_1,\xi_2)]^2
\right)^n
\right\rangle
= (-1)^n \, \sum_{\mu_1=1}^{2} \dots \sum_{\mu_n=1}^{2}
\frac{\partial^2}{\partial k_{\mu_1}^2} \dots
\frac{\partial^2}{\partial k_{\mu_1}^2}
\exp\left[ - \half \left( k_1^2 G_{1} + k_2^2 G_{2} \right) \right]
\right\vert_{k=0} \, .
\ee
Here, we have defined, for $\mu=1,2$,
\be\label{prop}
\left.
G_{\mu} = \nabla_{u,\mu}^2
  \,
\Bigl\langle \phi(0,0) \, \phi(\xi_1,\xi_2)  \Bigr\rangle \,
 \right\vert_{\xi=0} \, .
\ee
By using \eq{diff} for $n=2$ and translational invariance on the
torus, we find for the 2-loop expansion
\be
Z= {\rm const} \cdot \exp(-\sigma L_1 L_2) \cdot Z_1 \cdot
\left( 1 + \frac{R}{\sigma L_1L_2} \,
+ \dots \right) \, ,
\ee
with
\be\label{rdef}
R = \frac18 \, \left( 3 G_{1}^2 + 3G_{2}^2 + 2G_{1}G_{2} \right) \, .
\ee
 $R$ can be rewritten as
\be
R = \frac18 \, \left( 2 (G_{1}+G_{2})^2 + (G_{1}-G_{2})^2 \right) \, .
\ee
We now write $R$ as a sum of two parts,
\be
R = R_{\infty} + R_{\rm f} \, ,
\ee
with
\bea
R_{\infty} &=& \frac{1}{4}(G_{1}+G_{2})^2 \, , \\
R_{\rm f}  &=& \frac{1}{8}(G_{1}-G_{2})^2 \, .
\eea
With these definitions at hand, we now formulate our \\

\vspace{2mm}
\noindent
{\bf \underline{STATEMENT}}:

\vspace{2mm}

{\it
\noindent
At least for the four regularization  schemes \\[2mm]
{\rm
$\phantom{blob} \Lambda_1:\quad$ $\zeta$-function regularization, \\
$\phantom{blob} \Lambda_2:\quad$ point splitting  regularization, \\
$\phantom{blob} \Lambda_3:\quad$ continuous momentum space cutoff, \\
$\phantom{blob} \Lambda_4:\quad$ lattice cutoff,
}

\vspace{5mm}
\noindent
the following assertion holds:

\vspace{5mm}
\noindent
The coefficient $R$ of the 2-loop contribution to the
CWM partition function can be naturally split as
$R=R_{\infty} + R_{\rm f}$, and

\vspace{5mm}
\noindent
(S1) $R_{\infty}$ diverges when the cutoff parameter $a$ is sent to zero.
The singularity depends on the chosen regularization, but in all
regularization schemes it is ``shape independent''.
For any finite value of the cutoff, $R_{\infty}$ is
proportional to the area $L_1L_2$ of the rectangle, but does not depend
on the ratio $u=L_2/L_1$.

\vspace{5mm}
\noindent
(S2) $R_{\rm f}$ remains finite when the cutoff is removed:
\be
R_{\rm f} \longrightarrow
\frac18 \left[ 2 + \left( \frac{\pi}{3}
u E_2 \left( i u \right) - 1 \right)^2 \right] \, .
\ee

\noindent
$E_2$ denotes the first Eisenstein series:
\be
E_2(\tau)=1-24\sum_{n=1}^{\infty}\frac{n~ q^n}{1-q^n}~~,
\quad\quad q\equiv \exp(2\pi i \tau) \, .
\label{a10}
\ee
}
The calculation of $R$ by means of the $\zeta$-function regularization
can be found in~\cite{df}, while the analogous calculation with the
point-splitting procedure was presented in~\cite{thegang}. We refer to
the original papers for the details. Before we turn  to the proofs for
the  two schemes $\Lambda_3$ and $\Lambda_4$, let us state two immediate
 consequences of part $(S1)$ of the statement.  First, the fact that
$R_{\infty}$ depends on $L_1 \, L_2$ alone  (and not on the shape factor
$u$)  implies that  ratios of partition functions of interfaces with
different shapes but the same area stay finite when the cutoff is
removed. Second, this same fact allows to `absorb' the infinity  in a
`renormalization' of the interface tension $\sigma$  (see
also~\cite{thegang}).

\section{Proofs}

In this section we present the proofs of the above statement for the
cutoff schemes $\Lambda_3$ and $\Lambda_4$.  We return to dimensionful
variables in this section.

\subsection{Momentum Space Cutoff}

A very general cutoff procedure for fields on a torus
is to use the Fourier representation and then cut off
large momenta.
In such a scheme the regularized version of
$G_{\mu}$ can be written as
\be
G_{\mu}=
 - \sum_{p\neq 0}  \, \frac{p_{\mu}^2}{p^2} \, \chi(a^2p^2) \, .
\ee
The momenta are summed over
\be
p_{\mu} = \frac{2\pi}{L_{\mu}} \cdot \mbox{\ integer} \, ,
\ee
and $p^2=p_1^2+p_2^2$.
The cutoff function $\chi$ is assumed to have the property
$\chi(0)= 1$ and to drop sufficiently fast to zero when
the argument goes to infinity.
For technical reasons, we shall later
assume that there exists
a constant $c < 0$ such that
\be\label{assume}
|\chi(\rho)| \leq e^{c\rho} \, .
\ee
In order to remove the cutoff we have to send $a$ to zero.
Let us first study $G_{1}+G_{2}$:
\be
G_{1} + G_{2}=1 - \sum_{p} \chi(a^2 p^2) \, ,
\ee
where the sum is over {\em all} momenta (including 0).
We introduce a Fourier representation of the cutoff function
$\chi$:
\be
\chi(a^2 p^2) = \int d^2 x \, e^{ipx} \,  \tilde \chi_a(x) \, .
\ee
Here,
\be
\tilde \chi_a(x)
= \int \frac{d^2 p}{(2\pi)^2} \,  e^{-ipx} \, \chi(a^2 p^2) \, .
\ee
Now,
\be
\sum_p \chi(a^2 p^2) = \int d^2 x \, \tilde \chi_a(x) \, \sum_p e^{ipx} \, .
\ee
The sum over the momenta can be performed:
\be
\sum_p e^{ipx} = L_1 \sum_{n_1} \delta(x_1 - n_1 L_1) \,
                 L_2 \sum_{n_2} \delta(x_2 - n_2 L_2) \, .
\ee
Hence we have
\be\label{xxx}
\sum_p \chi(a^2 p^2) =
L_1 L_2 \sum_{n_1} \sum_{n_2} \tilde \chi_a(n_1 L_1,n_2 L_2) \, .
\ee
The Fourier transform of the cutoff function is
\begin{eqnarray}
\tilde \chi_a(x)
&=& \int \frac{d^2 p}{(2\pi)^2} \,  e^{-ipx} \, \chi(a^2 p^2) \nonumber \\
&=& \frac{1}{2\pi a^2} \int_0^{\infty} d\rho \,
\rho \, \chi(\rho^2) \, J_0(\rho |x| / a ) \, .
\end{eqnarray}
$J_0$ denotes the Bessel function of order 0.
Plugging this into eq.~(\ref{xxx}), we get
\be
\sum_p \chi(a^2 p^2) =
\frac{L_1 L_2}{2\pi a^2} \,
\sum_{n_1} \sum_{n_2}
\int_0^{\infty} d\rho \,
\rho \, \chi(\rho^2) \,
J_0 \left(\rho \sqrt{(n_1L_1)^2+(n_2L_2)^2} / a \right )
\ee
In the limit $a\rightarrow 0$ all contributions but the
$n_1=n_2=0$ term vanish. The argument is that $J_0$
oscillates wildly in the relevant interval, with the consequence
that for nonvanishing $n_i$ the integral vanishes
faster than $a^2$.
Since $J_0(0)=1$, we end up with
\be
\sum_p \chi(a^2 p^2) \rightarrow
\frac{L_1 L_2}{2 \pi a^2}
\int_0^{\infty} d\rho \, \rho \, \chi(\rho^2) =
\frac{L_1 L_2}{4 \pi a^2}
\int_0^{\infty} d\rho \, \chi(\rho) \, .
\ee
This completes the proof of the part $(S1)$ of the statement.

Let us now turn to the discussion of $G_{1}-G_{2}$.
In this case we use the Laplace transform of
$\chi$. For any function with
\be
|f(t)| \leq e^{ct}
\ee
there is a representation
\be
f(t)= \frac{1}{2\pi i}
\int_{c-i\infty}^{c+i\infty}
ds \, e^{ts} \, \hat f(s) \, .
\ee
For the cutoff function,
\be
\chi(a^2p^2)=
\frac{1}{2\pi i}
\int_{c-i\infty}^{c+i\infty}
ds \, \hat \chi(s) \, e^{s a^2 p^2} \, .
\ee
{}From the assumption \eq{assume} we infer that
we can choose $c < 0$. So the integration
is over a contour where the real part of $s$ is
always negative.
This allows for a representation of $G_{1} - G_{2}$ as follows:
\be\label{laplace}
G_{1}-G_{2}
=
- \frac{1}{2\pi i}
\int_{c-i\infty}^{c+i\infty} ds \, \hat \chi(s) \, F(s,a) \, ,
\ee
with
\be
F(s) \equiv - \sum_{p\neq 0}
\frac{p_1^2-p_2^2}{p^2} \,
e^{s a^2 p^2} \, .
\ee
Writing out explicitly the momenta, we get
\be
F(s,a)=-{\sum_{n_1,n_2}}^{\prime}
\frac{\frac{n_1^2}{L_1^2}-\frac{n_2^2}{L_2^2}}
{\frac{n_1^2}{L_1^2}+\frac{n_2^2}{L_2^2}}~
\exp \left[s \, 4\pi^2 a^2\left(\frac{n_1^2}{L_1^2}
+\frac{n_2^2}{L_2^2}\right)\right] \, .
\ee
Here, the prime on the summation symbol indicates that
$n_1=n_2=0$ is to be excluded.
Let us now separate the two terms in the numerator,
and let us call them $P_1$ and $P_2$.
\be
P_1=-\sum_{n_1,n_2} '\frac{\frac{n_1^2}{L_1^2}}
{\frac{n_1^2}{L_1^2}+\frac{n_2^2}{L_2^2}}~
\exp\left[ s \, 4\pi^2 a^2
\left(\frac{n_1^2}{L_1^2}+\frac{n_2^2}{L_2^2}\right)\right] \, ,
\ee
and
\be
P_2=\sum_{n_1,n_2} '\frac{\frac{n_2^2}{L_2^2}}
{\frac{n_1^2}{L_1^2}+\frac{n_2^2}{L_2^2}}~
\exp \left[s \, 4\pi^2 a^2
\left(\frac{n_1^2}{L_1^2}+\frac{n_2^2}{L_2^2}\right)\right] \, .
\ee
In $P_1$ we can safely take the limit $a\to 0$ in
the 2-direction, because, as a sum over the index $n_2$,  $P_1$ converges
absolutely. We find
\be\label{p1}
P_1 \hat = -\sum_{n_1,n_2} '\frac{\frac{n_1^2}{L_1^2}}
{\frac{n_1^2}{L_1^2}+\frac{n_2^2}{L_2^2}}~
\exp\left[ s \, 4\pi^2 a^2 \frac{n_1^2}{L_1^2} \right] \, .
\ee
Here we have introduced the symbol $\hat =$ which shall signal
equality in the limit $a \rightarrow 0$.
\Eq{p1} can be rewritten as follows:
\be
P_1 \hat = -2\sum_{n_1=1}^\infty \sum_{n_2=-\infty}^\infty
\frac{ u^2 \, n_1^2}
{n_2^2+ u^2 \, n_1^2}~
\exp\left[s\,4\pi^2a^2 \frac{n_1^2}{L_1^2} \right] \, .
\ee
The sum over $n_2$ can be done explicitly by using the
identity (see, e.g., text books on quantum statistics)
\be
\frac{b}{\pi}\sum_{n_2=-\infty}^\infty\frac{1}{n_2^2+b^2}=1+\frac{2}{e^{2\pi
b}-1} \quad \mbox{for} \quad b>0 \, .
\ee
We get
\be
P_1 \hat = -2\pi \, u  \left\{\sum_{n_1=1}^{\infty}
n_1\left(1+\frac{2}{e^{2\pi u n_1} -1}\right)
\exp\left[s \, 4\pi^2a^2 \frac{n_1^2}{L_1^2}\right]\right\} \, .
\ee
Let $q=e^{2\pi i \tau}=e^{-2\pi u }$. Then we have
\be
P_1 \hat =-2\pi u \sum_{n=1}^{\infty}\left(n
\exp\left[s 4\pi^2 a^2 \frac{n^2}{L_1^2} \right]\right)
-4\pi u \sum_{n=1}^{\infty}\left(\frac{nq^n}{1-q^n}
\exp\left[ s \, 4\pi^2a^2 \frac{n^2}{L_1^2} \right]\right) \, .
\ee
The second term is not singular in the limit $a\to 0$ while the first
term can be treated with the Euler-McLaurin summation formula:
\be\label{EULER}
\sum_{n=0}^k f(n)=\int_0^kf(x)\,dx +\frac12[f(0)+f(k)]
+\sum_{m=1}^\infty
(-1)^m\left[f^{(2m-1)}(k)-f^{(2m-1)}(0)\right]\frac{B_m}{2m!} \, ,
\ee
where the $B_m$ are the Bernoulli numbers.
Taking $f(n)=ne^{-bn^2}$ we get $f^{(2m-1)}(\infty)=0$ and
\be
\left.
f^{(2m-1)}(0)=-\frac1b \frac{d^{2m}}{dx^{2m}}e^{-bx^2} \right\vert_{x=0}=
b^{m-1}\frac{(-1)^m}{m!} \, ,
\ee
yielding
\be
\sum_{n=0}^\infty ne^{-bn^2}=\int_0^\infty dx \, xe^{-bx^2}
-\sum_{m=1}^\infty b^{m-1}\frac{B_m}{m!(2m)!} \, .
\ee
The last sum is an absolutely convergent series which is regular for
$b\to 0$.  The integral is elementary:
\be
\int_0^\infty xe^{-bx^2}\,dx=\frac1{2b} \, .
\ee
Taking $b= -s \, 4\pi^2a^2/L_1$, we get
\be
-2\pi u \sum_{n=1}^{\infty}\left(n
\exp\left[s 4\pi^2a^2 \frac{n^2}{R^2}\right]\right)\sim
\frac{L_1L_2}{s 4\pi a^2}
+\frac{\pi}{6} u +\cdots \, ,
\ee
where the omitted terms have higher powers of $a$.
The finite part of this sum is exactly what is needed to reconstruct
the Eisenstein function (once the limit $a\to 0$ is taken in the second
part of $P_1$) and we find:
\be
P_1 \hat = \frac{L_1L_2}{s 4\pi a^2}+\frac{\pi}{6} u E_2(iu) \, .
\ee
If we now study $P_2$ we see that we only have to exchange $L_1$ and $L_2$
and change the sign. In this way the divergent parts cancel out:
\be
F(s,a=0)= \frac{\pi}6 u E_2(iu)
-\frac{\pi}6 u^{-1} E_2(iu^{-1})
\ee
Using the identity
\be\label{tauschen}
E_2(-\frac{1}{\tau})=\tau^2 \, E_2(\tau)-i\frac{6\tau}{\pi}
\ee
we arrive at
\be
F(s,a=0) = \frac{\pi}{3} E_2(iu)-1 \, .
\label{r1}
\ee
Note that when we performed the limit $a \rightarrow 0 $, the  function
$F$ became also independent of $s$.  As a consequence, the Laplace
integral~\eq{laplace} then becomes just a representation of $\chi(0)$,
which  is one, and part $(S2)$ of the statement is proven.

\subsection{Lattice Cutoff}

In this case $G_{\mu}$ is defined as:
\be
G_{\mu}=
 - \sum_{p\neq 0}
\, \frac{\hat p_{\mu}^2}{\hat p_1^2 + \hat p_2^2} \, .
\ee
The momenta are summed over
\be
p_{\mu} = \frac{2\pi}{a l_{\mu}} \cdot n_{\mu}, \qquad n_{\mu}=0,
\dots, l_{\mu}-1 \, , \qquad L_{\mu}=a l_{\mu} \, ,
\ee
and
\be
a^2 \hat p_{\mu}^2 = 4 \, \sin^2 (a p_{\mu}) \, .
\ee
To remove the lattice cutoff one has to send $l_1$ and $l_2$ to
infinity, while keeping the ratio
\be
\frac{l_2}{l_1} = \frac{L_2}{L_1} \equiv u
\ee
fixed.
$G_1$ can be written as
\be
G_{1}=
 - \sum_{n_1=1}^{l_1-1} \left\{
   \sum_{n_2=0}^{l_2-1}
\, \frac{{\sin^2 \left(\frac{\pi n_1}{l_1} \right)}}
{{\sin^2 \left(\frac{\pi n_1}{l_1}\right)}
+{\sin^2 \left(\frac{\pi n_2}{l_2} \right)}}
\right\}
 \, .
\ee
The sum over $n_2$ can be done explicitly.
To this end we employ the identity
\be
\frac{\sinh^2(x/2)}{\sinh^2(x/2) + \sin^2(\omega/2) }
= \tanh(x/2) \sum_{n=-\infty}^{\infty}
\exp(-x|n| - i\omega n) \, ,
\label{a5}
\ee
which can, e.g., be derived by a series of elementary operations from
formula~3.613 of~\cite{gradstein}.
We make the identifications
\bea
\omega &=& 2\pi n_2/l_2 \, , \nonumber  \\
\sinh^2(x/2) &=& \sin^2(\pi n_1/l_1) \, .
\eea
Using that
\be
\sum_{n_2=0}^{l_2-1} \exp(-i \omega n) =
\sum_{n_2=0}^{l_2-1} \exp(-i \, 2\pi \, n \, n_2 / l_2 )
 = l_2 \sum_{k=-\infty}^{\infty} \,
 \delta(n - l_2 \, k) \, ,
\ee
we find after summing over $n$,
\be
\sum_{n_2=0}^{l_2-1}
\, \frac{{\sin^2\left(\frac{\pi n_1}{l_1} \right)}}
{{\sin^2 \left(\frac{\pi n_1}{l_1}\right)}
+{\sin^2 \left(\frac{\pi n_2}{l_2} \right)}}
= l_2 \, \tanh(x/2) \,
\left( 1+\frac{2 \, \exp(-l_2\,x)}{1-\exp(- l_2\,x)} \right) \, .
\ee
This allows us to identify the convergent and divergent parts in the limit
$l_1,l_2\to\infty$. We write
\be
G_{1} \equiv C_1 + D_1 \, ,
\ee
where
\be
D_1 = - l_2 \sum_{n_1= 1}^{l_1-1} \th(x/2) \, ,
\ee
and
\be
C_1 = - 2 \, l_2 \sum_{n_1= 1}^{l_1-1} \th(x/2) \,
\frac{\exp(- l_2\,x)}{1-\exp(-l_ 2\, x)} \, .
\ee
In all these sums the index $n_1$ is hidden in $x$.
$D_1$ can be written as
\be
D_1 =
- l_2 \sum_{n_1= 1}^{l_1-1} \frac{\sin \left(\frac{\pi n_1}{l_1}\right)}
{\sqrt{1+\sin^2 \left(\frac{\pi n_1}{l_1} \right)}} \, .
\ee
This sum can be evaluated by using the Euler-McLaurin formula~\eq{EULER}
and gives
\be
D_1 = -\frac{l_1l_2}{2}+ \frac{\pi}{6}  \, u + O \left( u / l_1^2 \right) \, .
\ee
In $C_2$ we can safely take the limit $l_1,l_2\to\infty$.
In this limit, we can write
$x=2\pi n_1/l_1+ O(1/l_1^3)$.
We can thus approximate $\th(x/2) \sim \pi n_1/l_1$, and
\be
\exp(- l_2 \,x)=\exp( -2\pi u  n_1 ) \,
\left( 1 + O \left(u/l_1^2\right) \right) \, .
\ee
Let us define as usual $q=e^{-2\pi u}$. Then
\be
C_1=- 4\pi \, u   \sum_{n=1}^\infty \frac{ n q^n}{1-q^n} \, .
\ee
Here we have taken into account that the two values $n_1$ and
$l_1-n_1$ are identified with the same value of $x$ thus giving a further
factor of 2.

Obviously, the corresponding results for $G_2$ can be obtained
just by letting $u \rightarrow u^{-1}$. Putting together the
various parts, we get
\bea
G_1 &=& D_1 + C_1 \rightarrow -\frac{l_1\, l_2}2 + \frac{\pi}{6} u E_2(iu) \, ,
 \nonumber \\
G_2 &=& D_2 + C_2 \rightarrow -\frac{l_1\, l_2}2 + \frac{\pi}{6} u^{-1}
E_2(iu^{-1}) \, .
\eea
Employing again~\eq{tauschen}, we now easily verify that
\bea
G_1 + G_2 &\rightarrow&  -(l_1\, l_2-1) \, , \nonumber \\
G_1 - G_2 &\rightarrow&  \frac{\pi}3 E_2(iu) - 1 \, ,
\label{r2}
\eea
which proves the assertions of the statement.

\section{Conclusions}
It could seem rather surprising that we find a universal result,
independent of the regularization scheme, for the finite part of $R$, since
it stems from a non-renormalizable operator. We think that the reason,
as it was already mentioned in~\cite{df} and~\cite{thegang}, is that the
result is completely fixed by symmetry requirements.
In particular, the modular invariance of the original interaction term (which
is a direct consequence of the fact that the model, due to the periodic
boundary conditions on $\phi$, is defined on a torus) and its scaling
behavior (which immediately follows from power counting) tells us that
$G_1-G_2$ must be a modular form of weight 2. The normalization coefficients
in front of the modular form then follow from the study of the
one-dimensional limit  $u\to\infty$, which was already discussed by Arvis
in~\cite{arvis}.

The contribution that we have found does not depend on the regularization
schemes that we have discussed because they all preserve the modular symmetry.
We are thus led to ask if there is some physical reason to choose among all the
possible schemes those which preserve this symmetry. We have no definite
answer on this point but it is tempting to conjecture that it is a consequence
of the fact that the Hamiltonians (\ref{Ag}) and (\ref{Ap}) are respectively
the first and the second order expansion of a bosonic string action, and that
the modular invariance is actually preserved at the quantum level in the
string model.
\section*{Acknowledgments}
We would like to thank F. Gliozzi, M. Hasenbusch, P. Provero, and S. Vinti
for many discussions. K.P. thanks G. M\"unster for interesting
discussions.


\end{document}